\documentclass[prl,twocolumn,superscriptaddress]{revtex4}

\usepackage{amsmath}
\usepackage{amsfonts}

\usepackage{graphicx}
\usepackage{times}

\begin{document}

%%%%%%%%%%%%

\title{Spin Order in Paired Quantum Hall States}

\author{Ivailo Dimov}
\affiliation{Department of Physics and Astronomy,
University of California, Los Angeles, CA 90095-1547}

\author{Bertrand I. Halperin}
\affiliation{Physics Deprtment, Harvard University, Cambridge, MA 02138}

\author{Chetan Nayak}
\affiliation{Microsoft Station Q, CNSI Building,
University of California, Santa Barbara, CA 93106-4030}

\begin{abstract}
We consider quantum Hall states
at even-denominator filling fractions, especially $\nu=5/2$,
in the limit of small Zeeman energy. Assuming that a
paired quantum Hall state forms, we study spin ordering
and its interplay with pairing. We give numerical evidence
that at $\nu = 5/2$ an incompressible
ground state will exhibit spontaneous ferromagnetism.
The Ginzburg-Landau theory for the spin degrees
of freedom of paired Hall states is a perturbed CP$^2$
model. We compute the coefficients in the
Ginzburg-Landau theory by a BCS-Stoner mean field
theory for coexisting order parameters, and show that
even if repulsion is smaller than that required for a Stoner
instability, ferromagnetic fluctuations can induce a 
partially or fully polarized superconducting state.
\end{abstract}

\maketitle

%%%%%%%%%%%%

\paragraph{Introduction.}
The spin ordering of the observed quantized Hall plateau
with $\sigma_{xy}=\frac{5}{2}\,\frac{e^2}{h}$ \cite{Willet87,Xia04}
has become a pressing issue due to its pertinence to
the identification of this state of matter as a potential platform
for topological quantum computation \cite{DasSarma07}.
Experimental \cite{Eisenstein88,Pan99} and numerical studies \cite{Morf98}
have not, thus far, settled the matter, although they are consistent
with a fully spin-polarized Moore-Read Pfaffian ground state
\cite{Moore91,Greiter92}.
In this paper, we revisit the spin-polarization of the ground state
at $\nu=5/2$ using (1) a variational Monte Carlo comparison of the energies of
polarized and unpolarized states, (2) a Ginzburg-Landau
effective field theory, and (3) a Fermi liquid calculation of the
magnetic instability of paired composite fermions. We find evidence that
it is polarized even if the Zeeman energy vanishes, $g{\mu_B}=0$.
Our analysis gives a simple physical picture for the energetics of various
states, drawing on similarities with ferromagnetic superconductors.

For large enough Zeeman energy, the ground state must
be fully polarized. However, the Zeeman energy in GaAs
2DEGs is small as a result of effective mass and
$g$-factor renormalization. Thus, the system is close to the limit
of strictly vanishing Zeeman energy, in which the Hamiltonian is symmetric under
the full SU(2) spin symmetry. At $\nu=1$ and $\nu=1/3$ in this limit,
the spins order ferromagnetically, thereby spontaneously
breaking this symmetry \cite{Sondhi93, Moon95}. However at $\nu = 5/2$,
an incompressible state is likely to exhibit pairing. It is thus
natural to ask if similar spin-ordering physics occurs at $\nu=5/2$, but from the
perspective that the ground state at this filling fraction is a 
spin-triplet paired state. Indeed, it is known \cite{Greiter92,Ho95} that
both the fully polarized Pfaffian and the unpolarized $(3,3,1)$ states
belong to the same family of triplet composite fermion pairs,
for which the wavefunction is $\Psi = \Psi_{\rm LJ}\Psi_{\vec{d}}$ with
$\Psi_{\rm LJ}=\prod_{j<k} (z_j - z_k)^m\prod_j e^{- |z_j|^2/4 }$
and
\begin{equation}
\label{eqn:Pfaff-d}
\Psi_{\vec{d}}  ~=~ {\rm Pf}\!\left(\frac{\vec{d}\cdot(i\vec\sigma\sigma_2)_{\alpha\beta}
|\alpha\rangle_j|\beta\rangle_k}{z_j - z_k }\right)~.
\end{equation}
with $\alpha, \beta = \uparrow,\downarrow$.
The complex unit vector $\vec{d}$ above is familiar from $^3$He physics.
For the fully polarized (along the $\hat{z}$-direction) Pfaffian state, 
$\vec{d}=-(\hat{x}+i\hat{y})/\sqrt{2}$, so the spin part of the pair 
wavefunction is $|\chi^s_{jk}\rangle = |\uparrow\rangle_j|\uparrow\rangle_k$
which has $S_z = 1$. The $(3,3,1)$ state corresponds to $\vec{d}=\hat{z}$,
for which each pair has $S_z=0$ and	
$|\chi^s_{jk}\rangle = |\uparrow\rangle_j|\downarrow\rangle_k+|\downarrow\rangle_j|\uparrow\rangle_k$ \cite{Ho95}.
In the language of $^3$He the $(3,3,1)$ state is 
therefore a {\it unitary} triplet paired state, while the Pfaffian
is a {\it non-unitary} triplet state \cite{Leggett75}. With this insight, it
was first observed by Ho \cite{Ho95} that one can 
obtain states in which the expectation value of the
spin of a pair, $\vec{F}=i\vec{d}\times\vec{d}^*$ has any value
$0\leq |\vec{F}| \leq1$.
Indeed, one can check that
the following:
\begin{equation}
\label{eq:d}
\vec{d} = \hat{z}(1-F^2)^{1/4}e^{2i\theta} - \frac{\hat{x}+i\hat{y}}{\sqrt{2}}
\left(1-\sqrt{1-F^2}\right)^{1/2}
\end{equation}
gives a {\it partially} polarized state with polarization magnitude $F$ for which
$|\chi^s_{jk}\rangle=\alpha(|\uparrow\rangle_j|\downarrow\rangle_k+
|\downarrow\rangle_j|\uparrow\rangle_k) + 
\beta|\uparrow\rangle_j|\uparrow\rangle_k$, where 
$\alpha=(1-F^2)^\frac{1}{4}\exp(2i\theta)$, 
and $\beta=(1-\sqrt{1-F^2})^\frac{1}{2}/\sqrt{2}$. 
A state with a polarization axis different from $\hat{z}$ can be obtained
by rotating $\vec{d}$.
%Because one can, in such manner, interpolate smoothly between the Pfaffian and
%$(3,3,1)$ states, it was conjectured by Ho\cite{Ho95} that there is a 
%crossover between those two states analogous to the $A$ to $A_1$ transition
%in $^3$He, and he proposed a model Hamiltonian that would connect the two states.
%However, later on Read and Rezayi \cite{Read96} showed that the model exibits a phase
%transition.

It is the purpose of this paper to analyze the energetics of spin 
for arbitrary triplet pairing, as well as the transitions between
unpolarized and partially- or fully-polarized states. 
We do this using several approaches.
First, we present a variational calculation in which we find that
the energy of the polarized Pfaffian is lower than that of the unpolarized 
$(3,3,1)$ at $\nu = 5/2$. This suggests that if the ground state
in the presence of Coulomb interaction is paired, it is fully or 
partially polarized. We then try to understand this result in
a larger context through the use of a Chern-Simons Ginzburg-Landau theory
for spinful electrons \cite{Lee90}, which we adapt to the case of a 
quantum Hall state of spin-1 bosons at even-denominator filling fraction.
We thus derive an effective field theory for the dynamics of the vector $\vec{d}$,
which turns out to be a perturbed CP$^2$ NL$\sigma$M 
model analogous to the O(3) NL$\sigma$M of quantum Hall ferromagnets
\cite{Sondhi93}. The SU(3) symmetry of the CP$^2$ model is lowered
to the physical SU(2) symmetry by the Zeeman coupling
$\tilde{g} = g{\mu^{}_B} B$ 
which couples to the composite pair
spin $\vec{F}$, and also by quadratic and quartic
spin-spin interactions,  $c_2$ and $u$. We analyze the resulting phase diagram 
as a function of $\tilde{g}$, $c_2$,
and $u$ and conclude that, for ${c_2}<0$, as expected for
a ferromagnetic pair-pair interaction, the system is either partially
or fully spin-polarized. If $u$ is sufficiently small (and, especially,
if it is negative), then the system is fully spin-polarized. The unpolarized
$(3,3,1)$ state only occurs in the event of antiferromagnetic pair-pair interactions.
Finally, we give a more microscopic
derivation of the effective field theory, thereby obtaining values for
$c_2$ and $u$, starting from a BCS-Stoner mean-field picture
of a triplet superconductor competing/cooperating with ferromagnetism.
It is important to include the magnetization as an independent
order parameter since the spins can order even if the composite
fermions do not pair, as in the case of compressible states 
\cite{HLR}. Since composite fermions have an enhanced
effective mass, this is a strong possibility and, indeed, this ordering
transition appears to have been observed in the compressible
state at $\nu=1/2$ \cite{Tracy07}. Moreover, the interplay between these two orders 
has recently garnered attention as a result of the discovery 
of ferromagnetic superconductors
such as ZrZn$_2$, UGe$_2$, and URhGe
\cite{UGe2,URhGe,ZrZn2,Powell03,Dahl07}, and because
such interplay can result in a transition between a unitary 
and a non-unitary triplet state. Except at the ordering transition,
the ferromagnetic order parameter can be integrated out, thereby
leading to the Ginzburg-Landau theory mentioned in the previous paragraph and
described below. However, the parameters $\tilde{g}$, $c_2$, and $u$
all receive important contributions from magnetic fluctuations, which we
compute. Our most interesting conclusion from 
this analysis is that even if short-range repulsion is insufficient to
trigger a Stoner instability, ferromagnetic fluctuations
can drive a transition to a partially polarized non-unitary state
once pairing is present.

\paragraph{Variational Monte Carlo calculation}
We can gain insight into which of the paired states (\ref{eqn:Pfaff-d})
are favored by variationally comparing the energies of the $(3,3,1)$ state and 
the Pfaffian. We have performed Variational Monte Carlo (VMC)
on the sphere for up to 60 electrons in both states in the spirit of
\cite{Morf87}. We have confirmed that at $\nu=1/2$
the energy per particle of Coulomb interaction is 
$E_{\rm Pf}/N = -0.457(2)$ in units of $e^2/\epsilon\ell_b$. However, we also 
find that the $(3,3,1)$ state is slightly {\it lower} in energy 
$E_{\rm 331}/N = -0.4634(5)$. This is still higher than the Composite Fermi Sea
(polarized or unpolarized \cite{Park98} ) in agreement with 
the absence of a plateau at $\nu=1/2$ \cite{HLR}. 
We analyze the $\nu=5/2$ case in the spirit of 
\cite{Park98} by mimicking the first Landau Level pseudopotentials of pure
Coulomb interaction with an effective interaction in the {\it lowest} Landau Level,
$V_{\rm eff}(r) = (e^2/\epsilon)(1/r + a_1e^{-\alpha_1r^2}+a_2r^2e^{-\alpha_2r^2})$.
In this case, we find that the Pfaffian is lower in energy than the $(3,3,1)$: 
$E_{\rm Pf}/N = -0.361(5)$, and $E_{331}/N = -0.331(5)$. 
This is in agreement with the existing numerical evidence
\cite{Morf98} that the ground state at $\nu = 5/2$ is spin-polarized. 
To decide if the lowest energy paired state is fully or partially polarized,
one would have to obtain the Coulomb energy of a partially polarized
state, which is hard to do variationally, because no efficient algorithms
for antisymmetrization exist.

\paragraph{CP$^2$ Ginzburg-Landau theory.}

The calculation of the previous paragraph indicated that
the ground state at $\nu=5/2$ is polarized. We now try to
understand this in the context of a Ginzburg-Landau effective field theory.
We begin with bosonic pairs with $e^* = 2$ at filling fraction ${\nu_b}=1/8$.
This corresponds to an electron filling fraction ${\nu_e}=1/2$.
(We ignore the filled $N=0$ Landau level
of the $\nu=5/2=2+\frac{1}{2}$
state and focus on the partially-filled $N=1$ Landau level):
\begin{multline}
\label{LG}
\mathcal{L} = \quad {\Psi^\dagger_i}(\partial_0 - 2i a_0)\Psi_i +
\frac{1}{2m^*}\left|( i\vec\partial + 2{\bf a} + 2{\bf A_{\rm ex}})\Psi_i \right|^2 \\
+ \frac{1}{2}v(2\Psi^\dagger_i \Psi_i - \bar\rho)^2
\:+ \: \frac{1}{4\pi\alpha}\epsilon^{\mu\nu\lambda}a_\mu\partial_\nu{a}_\lambda\\
+ \frac{1}{2}\int{d}^2r'
(\rho({\bf r})-\bar\rho)V({\bf r-r'})(\rho({\bf r'})-\bar\rho)\\
+ \frac{1}{2}g_{\rm eff}\mu^{}_B B\,{\Psi^\dagger_i}T^z_{ij}\Psi_j \:+\:
{c_2} \left({\Psi^\dagger_i} \vec{T}_{ij} {\Psi_j}\right)^2
+ u \left({\Psi^\dagger_i} \vec{T}_{ij} {\Psi_j}\right)^4.
\end{multline}
In Eq. \eqref{LG}, $m^*$ is the effective mass of a pair and 
at ${\nu_e} = 1/2$, $\alpha = 2$.
The bosonic order parameter ${\Psi_i}$, $i=0, \pm 1$ is essentially
$\vec{d}$: $\sqrt{\bar\rho/2}\,d_x = (\Psi_{-1}-\Psi_{1})/\sqrt{2}$,
$\sqrt{\bar\rho/2}\, d_y = (\Psi_1+\Psi_{-1})/i\sqrt{2}$, $\sqrt{\bar\rho/2}\, d_z = \Psi_0$. 
The advantage of using this basis is that the components
of the total spin $i\frac{\bar\rho}{2}\vec{d}\times\vec{d}^*=\Psi^\dagger_i
\vec{T}_{ij}\Psi_j$ become the generators of the usual spin-1 representation
of $SU(2)$, 
$T_x = \frac{1}{\sqrt{2}}
\left(
\begin{smallmatrix}
0 & 1 & 0 \\ 1 & 0 & 1 \\ 0 & 1 & 0
\end{smallmatrix}\right)
$,
$
T_y = \frac{1}{\sqrt{2}}
\left(
\begin{smallmatrix}
0 & -i & 0 \\ i & 0 & -i \\ 0 & i & 0
\end{smallmatrix}\right)
$,
$
T_z = 
\left(
\begin{smallmatrix}
 1 & 0 & 0 \\ 0 & 0 & 0 \\ 0 & 0 & -1
\end{smallmatrix}
\right)
$.
Therefore the top and bottom components of $\Psi_i$ represent Pfaffian states
along the $S_z\pm1$ direction while the middle component is a $(3,3,1)$ state
with $S_z=0$. In addition to the familiar  Chern-Simons
Ginzburg-Landau and Coulomb interaction terms \cite{Zhang89}, the
last line of \eqref{LG} contains a Zeeman energy term coupling to the
pair spin, as well as quadratic and quartic spin-spin interaction
terms, $c_2$ and $u$ respectively. These couplings can,
in principle, be derived from the underlying
composite fermion theory from which \eqref{LG} emerges at
length scales longer than the pair size. This is done in a simple
model below. The Zeeman coupling $g_{\rm eff}{\mu_B}$ is an effective
parameter after the fermions are integrated out.
The Coulomb exchange interaction
between fermions induces a ferromagnetic interaction between pairs.
However, in a Stoner picture for itinerant fermion ferromagnetism,
exchange must compete with kinetic energy. This competition is
reflected in $c_2$, as we see by explicit calculation
later. If ferromagnetic exchange dominates, $c_2 < 0$ and a fully
polarized Pfaffian or a partially polarized state becomes the ground state,
but for now we consider both signs.
Finally, in the description of spin-$1$ atoms
in an optical trap (`spinor condensates'), the quartic coupling, $u$, 
would be negligible since the probability for 4 bosons to meet at a point 
is extremely small at low density \cite{Ho98}.
In a system of weakly-bound BCS-like pairs, however, such
a term need not be small since the pair size is comparable to
the spacing between pairs. This Ginzburg-Landau theory (\ref{LG}) is valid at
energies below the pairing gap $\Delta_0$ to neutral fermionic excitations.
In this regime, the fermions may be integrated out
so long as no vortices are present.
Later, we will do this explicitly in a simple model
in order to derive the Ginzburg-Landau effective field theory.
When vortices are present, we must be more careful, since
there will be fermionic zero modes which
are crucial for the non-Abelian braiding statistics of vortices
\cite{Nayak96c,Read96,Fradkin98,Read00,Ivanov01}).

When $\Psi_i$ condenses, we can write it as 
$\Psi_i = \sqrt{\bar\rho/2}\,e^{2i\theta}\xi_i$ with
${\bar \xi_i} \xi^{}_i = 1$. Since $\xi_i$, which transforms as
a vector under spin rotations, is complex and of unit magnitude
it takes values in CP$^2$. Substituting $\Psi_i$ into \eqref{LG}
one can see that kinetic energy will be relieved if charge fluctuations
$J^c_\mu = 2\partial_\mu\theta$ and spin fluctuation
$J^S_\mu = \bar\xi_i\partial_\mu\xi_i$ cancel. 
But because both currents are charged due to the Chern-Simons
gauge field, there is a Coulomb self-energy cost associated with 
both vortex and Skyrmion excitations.
This energy cost favors large skyrmions and competes with
the Zeeman energy, which favors small skyrmions.
We follow the steps outlined in 
Kane and Lee \cite{Lee90}\footnote{Also see the discussion
following Eq. (192) in \cite{Moon95}},
and obtain a perturbed CP$^2$ model for the 
$\xi_i$ variables alone, which is a generalization of
the perturbed NL$\sigma$M of quantum Hall ferromagnets
\cite{Sondhi93}:
\begin{multline}
\label{eqn:CP2-LG}
\mathcal{L}_{\rm eff} = \rho \,\bar{\xi}_i {\partial_0}{\xi}_i +
 \frac{1}{2}K({\partial_i}{\bar{\xi}_i}{\partial_i}{{\xi}_i}
+ {({\bar{\xi}_i}{\partial{\xi}_i})^2}) + \mathcal{L}_{\rm Hopf}\\ +
g_\text{eff} {\mu_B}B{\bar\rho}\left({|\xi_1|^2 - |\xi_{-1}|^2}\right) + 
{\tilde{c}_2}{\left({\bar{\xi}_i}\vec{F}_{ij}{\xi_j}\right)^2} +
{\tilde{u}}{\left({\bar{\xi}_i}\vec{F}_{ij}{\xi_j}\right)^4}\\ + 
\frac{1}{8\alpha^2}\int{d}^2r'Q_{Sk}({\bf r}) V({\bf r-r'}) Q_{Sk}({\bf r'})
\end{multline}
In the above $\tilde{c}_2 = c_2\bar\rho^2$ and $\tilde{u} = u\bar\rho^4$,
but for simplicity from now on we will omit the tildes.
Here, $Q_{Sk} = (-i/2\pi)\epsilon^{\mu\nu}\partial_\mu\bar\xi_i\partial_\nu\xi_i$, is the $\text{CP}^2$ Skyrmion charge density.
A charge-one  $\text{CP}^2$ Skyrmion carries electrical charge $e/4$, just as a vortex.
A conventional Skyrmion texture in the magnetization vector
$n_i = i\epsilon_{ijk}\xi_j \times {\bar\xi}_k$
has $\text{CP}^2$ Skyrmion charge $2$.
The Hopf term in the first line of \eqref{eqn:CP2-LG} gives the Abelian 
part of the Skyrmion statistics.

\begin{figure}[htb]
\centerline{\includegraphics[height=1.5in]{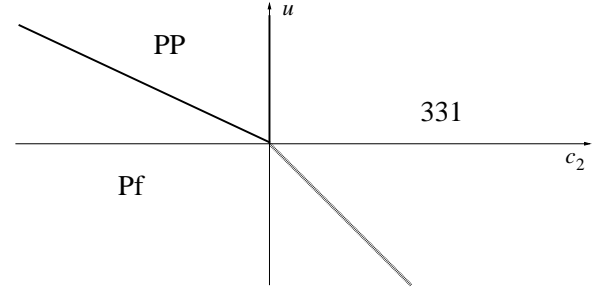}}
%\vskip 0.5cm
\caption{Phase diagram for zero Zeeman coupling as a function of $\tilde{c}_2$ and 
$\tilde{u}$, as explained in the text}
\label{fig:Pfaff-phase-diag}
\end{figure}

\paragraph{Phase Diagram.}
Let us consider the ground state of (\ref{eqn:CP2-LG}).
The Hopf term is unimportant for energetics and so is
the Coulomb energy of charged excitations.
For $g=u=c_2=0$, the system is at a (multi-)critical point controlled
by the CP$^2$ model. At this critical point,
the Pfaffian, the $(3,3,1)$, and all states interpolating between them have
the same energy. The phase diagram for $g = 0$, and
general $\tilde{c}_2$, $\tilde{u}$ has the form
depicted in figure \ref{fig:Pfaff-phase-diag}.
For ${c_2},u>0$, the system is in the $(3,3,1)$ phase.
For $\tilde{u}<0$, there is a first-order phase transition
at $\tilde{c}_2=-\tilde{u}>0$ from the $(3,3,1)$ state to
the fully-polarized Pfaffian state. This is both a topological
phase transition and a conventional (${\xi_y}\rightarrow-{\xi_y}$)
$Z_2$ symmetry-breaking
transition. For $\tilde{u}>0$, there is
a second-order phase transition at ${c_2}=0$ from the
$(3,3,1)$ phase to a partially-polarized (PP) state, which is
a conventional $Z_2$ symmetry-breaking transition.
In a wedge of the phase diagram between
the lines $-\tilde{c}_2 = 2\tilde{u}$ and $\tilde{c}_2=0$ with $\tilde{u}>0$,
each pair has ${F^2}= -\tilde{c}_2/2\tilde{u}\leq1$.
At $-\tilde{c}_2 = 2\tilde{u}$, $\tilde{u}>0$ the system
becomes fully-polarized. This is a second-order phase transition
at the mean-field level, but there is no symmetry distinction
between the partially and fully-polarized states. However,
when we take into account the underlying fermions, there
will be a topological phase transition between Abelian and non-Abelian states.
In general, this transition will not occur at $-\tilde{c}_2 = 2\tilde{u}$
but, instead, before the system becomes fully-polarized \cite{Read00}.
This will be discussed further elsewhere \cite{NayakDimov07}.
All of these phases have gapless spin excitations
which are the Goldstone modes of spontaneously-broken $SU(2)$. 
Finally, turning on the $SU(2)$ 
symmetry-breaking perturbation $g$ always induces non-zero
magnetization. For $g>2({c_2}+2u)$, the system is fully-polarized.
We now turn to a more microscopic calculation of the parameters
$c_2$ and $u$.

\paragraph{BCS-Stoner calculation of $c_2$, $u$.}
Following Greiter et.al. \cite{Greiter92}, by using flux attachment 
we can consider electrons at half filling as composite Fermions (CFs)
which would be free if the Chern-Simons gauge field is replaced by its mean
field value. The CFs would then form a Fermi sea \cite{HLR} and one can 
ask what the effects of the Chern-Simons gauge field fluctuations would be
on the Cooper pair channel.  Greiter {\it et al.}
showed that the Cooper channel contribution coming from this gauge interaction is 
triplet and the usual BCS analysis favors a $p+ip$ superconductor.
However, in the absence of pairing, CF effective mass renormalization \cite{HLR}
and Coulomb repulsion would also favor ferromagnetism. Therefore,
as a starting point, we assume the following BCS-Stoner reduced Hamiltonian:
\begin{multline}
H = \sum_k \frac{k^2}{2m^*}c^\dagger_{k\alpha}c_{k\alpha}
+ \vec{M}\cdot c^\dagger_{k\alpha} \vec{\sigma}_{\alpha\beta}c_{k\beta}
\\ + 
\vec{\Delta}^*_k\cdot{c}_{k\alpha}(i\vec{\sigma}\sigma_2)_{\alpha\beta}c_{-k\beta}
+ {\rm h.c.} 
\label{eq:reduced-bcs-stoner}
\end{multline}
For simplicity, we assume short range repulsion
$\vec{M}= U \sum_k c^\dagger_{k\alpha} 
\vec{\sigma}_{\alpha\beta}c_{k\beta}$. 
The role played by the Chern-Simons gauge field
is apparent only through the interaction
$V_{kk'}=\pi\,\vec{k}\times\vec{k'}/|\vec{k}\times\vec{k'}|^2$ which
enters the self-consistency condition, $\vec{\Delta}_{k} =
V_{kk'}\left\langle (i\vec{\sigma}\sigma_2)_{ab} {c}_{k'a}c_{-k'b}\right\rangle$.
We note that Eq. \ref{eq:reduced-bcs-stoner} represents, in principle, a more general
class of states than (\ref{eqn:Pfaff-d}) since it is not assumed that $\vec{F}=\vec{M}$.

The spectrum of Bogoliubov-de Gennes quasiparticles resulting from 
\eqref{eq:reduced-bcs-stoner} is:
\begin{multline}
E^2_{k\pm} = \tilde\epsilon^2_k +\Delta^2_{k}+M^2\\
\pm\sqrt{|\vec\Delta_{k}^*\times\vec\Delta_{k} + 2i\tilde\epsilon_k\vec{M}_k|^2+
4|\vec{M}\cdot\vec\Delta_k|^2}
\label{eq:energy}
\end{multline}
where $\tilde\epsilon_k = \epsilon_k - \mu$ and we assume that $\vec{\Delta}_k$ has
the chiral p-wave form \cite{Greiter92}
$\vec{\Delta}_k={\Delta_0}\vec{d}({k_x} + i {k_y})/{k_F}$ if $k<k_F$ and
$\vec{\Delta}_k={\Delta_0}\vec{d} k_F/({k_x} - i {k_y})$, if $k>k_F$,
where $\vec{d}$ is a complex unit vector, as before. We
integrate out the fermions to obtain the effective potential:
\begin{multline}
V_{\rm eff} = \int_k \sum_{\sigma=\pm}(\tilde\epsilon_k - E^{}_{k\sigma})
+\int_{kk'}\vec\Delta^*_kV^{-1}_{k',k}\vec\Delta_k + \frac{M^2}{2U}
\label{eq:veff}
\end{multline}
We take $\Delta_0$ fixed and expand to
fourth order in $M$, and $F = i\vec{d}\times\vec{d}^*$,
thereby expanding about the $(3,3,1)$ state. We can thereby study the tendency
of the system to develop a magnetization, although we will not be able
to access the fully-polarized limit in this approximation.
We obtain terms coupling the two order parameters:
\begin{multline}
\label{eqn:V-eff-expand}
V_{\rm eff}(\vec{M},\vec{F},\vec{\xi}) = {\alpha_2} F^2
+ {\alpha_4} F^4 +  {\gamma_1}\vec{F}\cdot\vec{M} + {\gamma_3} F^2 \vec{F}\cdot\vec{M}\\
+ M_iR_{ij}M_j\vec{F}\cdot\vec{M} + \chi^{-1}M^2 + B_{ij}{M_i} {M_j} \\
+ u_a M^4 + u_b M^2 |\vec{d}\cdot\vec{M}|^2 + u_c |\vec{d}\cdot\vec{M}|^4
\end{multline}
where ${\alpha_2}=m^*\pi\Delta^2_{0}$, ${\alpha_4}=m^*\pi\Delta^2_{0}/6$,
${\gamma_1}=2m^*\pi\epsilon_F\eta^2$, 
${\gamma_3}=-2m^*\pi\epsilon_F\eta^2/7$, $u_m = 3m^*\pi/2\epsilon^2_F\eta^2$,
$u_{md} = 4m^*\pi/3\epsilon^2_F\eta^2$, and $u_d = 8m^*\pi/3\epsilon^2_F\eta^2$, 
with $\eta\equiv\Delta_0/\epsilon_F$ assumed to be small 
but non-zero (since the effective expansion parameter is ${M_0}/\Delta_0$).
We also have $R_{ij} = r_m\delta_{ij} + r_d d_id^*_j$ and 
$B_{ij} = {B_d} d_id^*_j + {B_F} F_i F_j$, with $r_m = -8m^*\pi/\epsilon_F$,
$r_d = -40m^*\pi/7$, $A = 1/U - 4m^*\pi$, ${B_d} = (4m^*\pi(1 + F^2/3))$,
and ${B_F} = 2m^*\pi/3$. A similar result has been found in \cite{Dahl07}
in the limit that
both $\Delta_0$ and $M$ are small (which is different from our limit of small
$\vec{F}$, $\vec{M}$ but finite $\Delta_0$).

The coupling between magnetism and superconductivity
enhances the tendency to magnetism. $\chi^{-1}>0$
would disfavor a magnetic moment in the absence of pairing;
${\alpha_2}, {\alpha_4}>0$ would favor unitary ground states. However,
the coupling between magnetism and triplet superconductivity
can lead to a non-zero moment and a non-unitary order parameter
even when $\chi^{-1},{\alpha_2}, {\alpha_4}>0$.
The condition is essentially that the smallest eigenvalue of the
matrix ${\partial^2}V_{\rm eff}/\partial {X_i}\partial{X_j}$, with
$X=(\vec{M},\vec{F})$, become negative. This occurs when
${\alpha_2}\chi^{-1} < {\gamma_1^2}/4$ or, equivalently,
using the expressions after (\ref{eqn:V-eff-expand}),
$\frac{1}{U}-(4+\eta^2)m^*\pi<0$.

If we diagonalize the quadratic terms and integrate out the
fields which correspond to the positive eigenvalues
of ${\partial^2}V_{\rm eff}/\partial {X_i}\partial{X_j}$,
we obtain an effective action of the form of (\ref{eqn:CP2-LG})
with $c_2$ and $u$ given by:
\begin{equation}
c_2 = \left(\frac{1}{U}-(4+\eta^2)m^*\pi\right)\!\frac{\Delta_0^2}{2} \,,
\hskip 0.3 cm u = \frac{3\pi m^* \epsilon^4_F}{2\Delta_0^2}.
\label{eq:c2-u}
\end{equation}
As $U$ is increased from zero, the system undergoes
a second-order phase transition
from the $(3,3,1)$ state to a partially-polarized state.
The expressions (\ref{eq:c2-u}) are only valid for small $\eta$,
but for larger $\eta$ a second transition will occur to the
fully-polarized Pfaffian state \cite{NayakDimov07}.
This is likely to be the physically relevant regime
for the $\nu=5/2$ state, where there is only one energy ${e^2}/\epsilon \ell_0$,
which sets the scale for both ${\Delta_0}$ and ${\epsilon_F}$.

\paragraph{Discussion}
From the results described above, we see that if the $\nu=5/2$ quantum
Hall state is a spin-triplet paired state, then it will be polarized
in the limit of sufficiently strong ferromagnetic interactions.
Whether or not this occurs and whether it is partially or fully polarized
depends on the strength of the short range replulsion relative to the
effective fermion mass. Large repulsion would favor full polarization, while
repulsion comparable to the effective mass would favor partial polarization
even if lower than the Stoner critical value.
While we do believe our mean-field BCS-Stoner model captures 
the essential physics, one should be careful before comparing with
experiments, because we have not taken into account, for example, effective
mass divergences at the Fermi surface, which are known to arise at
$\nu=1/2$ \cite{HLR, Read98}. Another important issue
concerns the identification of the excitations in the various
partial and fully polarized states. One crucial question that
begs an answer is: do the excitations carry non-Abelian statistics? We discuss
this elsewhere 	\cite{NayakDimov07}.

\paragraph{Acknowledgements}
We would like to thank S. Das Sarma, E. Demler, J. Eisenstein, K. Yang, R. Morf,
and E. Rezayi for discussions.
This research has been supported by the NSF under grants
DMR-0411800 and DMR-0541988; by the ARO under grant W911NF-04-1-0236;
and by the Microsoft Corporation.

\vskip -0.5cm

%\bibliography{../corr}
%\bibliographystyle{prsty}

\end{document}